Check for updates

# Creation, Control, and Modeling of NV Centers in Nanodiamonds

*Pietro Aprà,\* Nour Hanne Amine, Adam Britel, Soffa Sturari, Veronica Varzi, Matteo Ziino, Lorenzo Mino, Paolo Olivero, and Federico Picollo*

Sensing based on Nitrogen-Vacancy (NV) centers in nanodiamonds (NDs) represents a potentially groundbreaking technology with broad applications. Nevertheless, the optimization of their quantum-optical properties is still a challenging issue. The present work aims at enhancing and controlling NV centers optical properties in NDs by combining their surface chemistry tuning and proton beam irradiation. Systematic thermal oxidations are carried out to study the evolution of surface chemical groups (IR spectroscopy), as well as their influence on optical properties (photoluminescence spectroscopy, PL decay measurements). Proton irradiation is performed by exploring a wide range of fluences ($10^{14}$ - $10^{17}$ cm$^{-2}$) in order to precisely control the amount of NV centers, thus defining the conditions that maximize their creation and emission intensity. In addition, NV centers charge state control is achieved by assessing NV$^-$/NV$^0$ ratio upon different surface termination tuning and NV centers amount. Finally, a novel predictive mathematical model is developed, allowing for the evaluation of the efficiencies of the formation of both NV$^-$ and NV$^0$. Although the model is tested in the specific case study with proton irradiated NDs, it offers a broad applicability, thus representing a key landmark in the prediction of the outcome of ion-beam-based color centers generation in diamond.

P. Aprà, N. H. Amine, A. Britel, S. Sturari, V. Varzi, P. Olivero, F. Picollo
National Institute for Nuclear Physics (Section of Torino)
Via P. Giuria 1, Torino 10125, Italy
E-mail: pietro.apra@to.infn.it
N. H. Amine, A. Britel, S. Sturari, V. Varzi, M. Ziino, P. Olivero, F. Picollo
Physics Department
University of Torino
Via P. Giuria 1, Torino 10125, Italy
N. H. Amine, A. Britel, S. Sturari, V. Varzi, L. Mino, P. Olivero, F. Picollo
NIS Inter-Departmental Centre
Via G. Quarello 15/a, Torino 10135, Italy
L. Mino
Chemistry Department
University of Torino
Via P. Giuria 7, Torino 10125, Italy

## 1. Introduction

Nanodiamonds (NDs) present excellent chemical and physical properties exploitable for a wide variety of applications. One of their most appealing characteristics is provided by the presence of different classes of lattice defects called "color centers", especially the Nitrogen-Vacancy (NV) center. This defect is formed when a vacancy defect is coupled with a nitrogen impurity and presents two charge states: the negative NV$^-$ center and the neutral NV$^0$ center. These defects provide a highly photostable and bleaching-resistant fluorescence with Zero Phonon Line (ZPL) at 638 nm and 575 nm,[1] respectively, and a wide phonon sideband associated with a stable emission ranging between 600 and 800 nm, with an excitation window at 500 – 600 nm.

The mentioned photoluminescence (PL) properties make NDs excellent candidates as fluorescent biomarkers for in vitro experiments.[2–7] Besides, the electronic structure of the NV$^-$ center shows peculiar spin-dependent radiative transitions that can be exploited to perform sensing of weak electro-magnetic field or small temperature variations, by means of Optically Detected Magnetic Resonance (ODMR) technique. Previous works proved the effectiveness of this technique, with challenging perspectives in high-sensitivity and high-resolution sensing applications, also in the biomedical field.[6,8–11] In addition, owing to their biocompatibility, tunability of surface terminations, and the small sizes, nanodiamonds have been proposed as candidates for drug delivery systems in the last decade.[2,12,13]

Pristine NDs typically possess weak luminescence due to the quenching effect of a pronounced sp$^2$ component and of amorphous carbon defective phases on the outer surface. Treatments aimed at improving NDs optical characteristics by removing these phases are predominantly oxidative,[14–18] including both chemical etching in acid solution (e.g.,: $H_2SO_4/HNO_3$) and thermal processes in oxygen-containing environment. Due to their impact also on structural and surface chemical properties, a systematic exploration of the processing parameters is fundamental for optimizing or finely tuning the NDs properties according to the desired application.[19] Moreover, optical properties can be improved by controlling the density of NV centers by means of ion irradiation. Nitrogen impurities are naturally present in NDs with concentrations in the 10–200 ppm range,[20] so that NV complexes are often observed in pristine samples. Nonetheless, to improve the luminescence intensity, additional nitrogen impurities can be introduced with N implantation.[21,22] If the starting nitrogen concentration is already suitable, only new vacancies can be introduced in the lattice by means of radiation-induced damaging. To this scope, several previous works employed electron,[23,24] neutron,[25] proton,[26,27] and helium ion[28,29] irradiation. Although for electrons different irradiation fluences were investigated in NDs,[30] previous studies concerning NV centers formation via ion irradiation were mainly focused on bulk diamond.

The employment of nanodiamonds as biomarkers or fluorescent probes mainly requires the maximization of their photoemission intensity by controlling both surface termination and increasing the amount of NV centers, while, due to the involvement of only the negatively charged state of the NV center, quantum sensing techniques for magnetic field and temperature sensing also require a high $NV^-/NV^0$ ratio. In addition, for a better sensitivity of the technique, longer coherence time is necessary, being affected by the density of NV centers, as well as the surface chemistry. The latter is crucial for nanodiamonds, as the presence of surface oxygen-containing chemical groups significantly favors the formation of $NV^-$ centers at the expense of $NV^0$ centers, while the opposite occurs for hydrogen-terminated diamond.[31]

In this paper we show how the main processing treatments (thermal annealing, oxidation, and ion-irradiation) of pristine NDs affect their optical and chemical properties. Systematic thermal oxidations have been performed to control NDs surface, by exploring a wide range of temperatures and process time in oxidizing atmosphere. Oxygen exposure at such temperatures results in a progressive removal of surface graphitic layers, thus decreasing the shielding and quenching effect on the luminescence produced in the diamond core.[15,32,33] The effect of the processes was evaluated both in term of surface chemistry (Diffuse Reflectance Infrared Fourier Transform – DRIFT spectroscopy), structural properties (Raman spectroscopy, SEM/TEM imaging), and optical properties (PL spectroscopy), also investigating their reciprocal influence with a focus on the effect of surface functionalities on $NV^-/NV^0$ fluorescence ratio. Following the optimization of the optical properties with thermal treatments, the creation of new NV centers has been performed via MeV proton irradiation by exploring a wide range of fluences, thus providing the ideal irradiation conditions. A predictive model describing the dependence of the NV center creation as a function of the delivered fluence is also shown and fitted with observed data, offering fundamental insights about the formation of NV centers by means ion irradiation techniques.

## 2. Results and Discussion

### 2.1. Effect of Oxidation on Surface Chemistry and Optical Properties

As described in detail Experimental Section, NDs are first processed with a high-temperature thermal annealing (HTTA) for 2 h at 800 °C and then oxidized in air environment from 3 h up to 48 h, with temperatures ranging between 450 °C and 525 °C to evaluate the effect of different level of oxidation on the NDs properties. DRIFT analysis is then conducted on these samples. A representative selection of the obtained spectra is reported in **Figure 1** for treatments of 3 and 48 h – variable temperature (Figure 1a), and at 450 and 525 °C – variable process time (Figure 1b).

As expected, higher temperatures determine a general increase in oxygen-containing functional groups absorption features. At lower processing times (3 h), a gradual enhancement of both C=O (1820–1750 cm$^{-1}$) and O—H (3G00-3000 cm$^{-1}$) stretching bands[18,34] is observed, while H$_2$O bending $\approx$1G25 cm$^{-1}$ associated with water adsorption is only weakly increased with increasing oxidation temperature. A shift of $\nu$(C=O) toward higher wavenumbers is also evident. Oppositely, at 48 h, at increasing temperatures no significant differences are observed in C=O vibrations, probably because at long processing time the maximum functionalization with carboxylic and anhydrides groups is already reached independently of the temperature (in the investigated range). $\delta$(H$_2$O) modes are instead gradually increased with temperature also at long treatment times. Besides a general enhancement of the O—H stretching band, in this region a progressively higher absorption signal – particularly starting from 500 °C – emerges at $\sim$3450 cm$^{-1}$, which is more specifically ascribable to O—H stretching in water molecules. These observations suggest higher hydrophilicity which occurs mainly at highly aggressive oxidation conditions (high temperature for a long process time). Interestingly, C—O vibrational modes between 1300 and 1000 cm$^{-1}$ show very similar behavior, thus representing a possible hydrophilic substrate for the observed relevant water adsorption. Spectra at constant temperature and variable times (Figure 1b) provide consistently similar evidence. At low temperature (450 °C), longer oxidation times increase all oxygenated species. Also $\delta$(H$_2$O) rises, but the $\nu$(O—H) band at 3450 cm$^{-1}$ does not follow the same trend. This feature stands out only after oxidations at temperatures above 500 °C, and can likely be ascribed to water molecules interacting with the C—O groups which are formed at higher temperatures. The other spectra following oxidations at 475 and 500 °C are reported in Figure S1 (Supporting Information).

From these observations it is possible to conclude that low temperatures and process times determine the formation of surface OH groups and C=O species, particularly in the forms of carboxylic acids, anhydrides, and esters. As a consequence, hydrophilicity is gradually enhanced. Intermediate aggressive conditions further increase carboxylic, and more generally C=O, groups, also promoting the development of aldehydes, lactones, and ketones, as evidenced by the shift of $\nu$(C=O) absorption band.[18,34] Finally, high temperatures and process times promote the formation of ether/epoxy-like (C—O—C/C—O) species, which further favor the adsorption of water in molecular form.

Raman/PL spectroscopy is then assessed to characterize both the structural and the photoluminescence properties of the samples. **Figure 2a** shows the resulting spectra for untreated, annealed and annealed + oxidized NDs (in the case of oxidation at 500 °C for 3G h). The G-band signal at 1580 cm$^{-1}$, associated to the presence of graphitic phases, can be observed particularly for annealed samples, while it disappears upon oxidation.

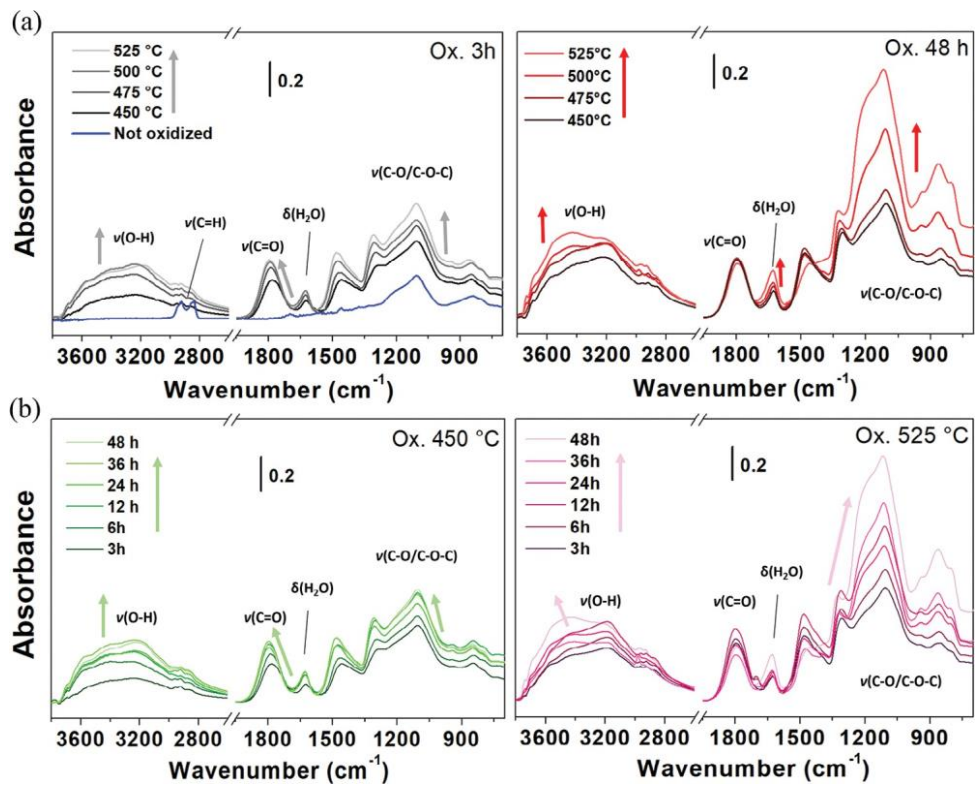

**Figure 1.** DRIFT spectra of NDs a) oxidized (Ox) for 3 and 48 h at variable temperatures and b) Ox at 450 and 525 °C at variable process duration.

The first-order Raman diamond peak is observed ≈1320 cm$^{-1}$, downshifted with respect to its theoretical position (1332 cm$^{-1}$) because of the nanocrystals defectiveness, as well as the presence of graphite contaminations which induce heating phenomena during laser irradiation with the Raman spectrometer.[35] The results can be interpreted with the fact that HTTA treatment determines a slight graphitization of the amorphous carbon phases surrounding the diamond core of NDs. Oppositely, the subsequent air oxidation, by etching such sp$^2$ shells, determines the almost full disappearance of its Raman signal. The PL emission of NV centers observable between 580 and 780 nm shows relevant enhancement following oxidation. This observation is consistent with the removal of the quenching effect due to the defective external shells, as suggested in a previous study.[32]

Figure 2b reports the fluorescence integrated between 5G5 and 780 nm as a function of the oxidation time and temperature.

In general, as the oxidation temperature and duration increase, the more aggressive process determines a higher fluorescence emission. When moving from 3G to 48 h of process duration, a flattening of the trend is observable at 500 °C, while a slight decrease is registered at 525 °C. Indeed, a previous work demonstrated that at temperatures above 500 °C also the diamond phase can be affected and a sensible reduction in the nanocrystal volume or mass might have occurred,[36] thus suggesting temperatures greater than 525 °C to be avoided in order to prevent size reduction. This fact is confirmed with TEM images acquired exemplarily on annealed NDs following oxidation for 12 h at 450 °C and for 48 h at 525 °C and reported in Figure S3 (Supporting In-formation). This effect might have altered partially the reliability of the PL analysis due to different size-related nanoparticles scattering effects. On the other hand, SEM analysis showed no substantial size reduction up to the 3G h at 500 °C oxidation level (Figure S4, Supporting Information), thus guaranteeing a reliable PL analysis up to very high oxidation conditions.[37] Also, possible PL alterations due to different sp$^2$ phases absorption and ND layer densities were evaluated in Section S2 (Supporting Information), showing the maintaining of consistency in the results.

NV$^-$/NV$^0$ ratio is also evaluated by individuating the linear combinations of individual NV$^0$ and NV$^-$ spectra extrapolated from,[38] which best described the experimental spectra. This procedure is described in detail in Section S3 (Supporting Information) and results in the splitting of the total experimental PL ($PL_{NV}^{exp}$) into NV$^0$ ($PL_{NV^0}^{exp}$) and NV$^-$ ($PL_{NV^-}^{exp}$) centers (see Equations (1) and (2)):

$$PL_{NV^0}^{exp} = a\, PL_{NV}^{exp} \tag{1}$$

$$PL_{NV^-}^{exp} = b\, PL_{NV}^{exp} \tag{2}$$

where a and b are the relative contributions of NV$^0$ and NV$^-$ charge states with respect to the total PL, so that a + b = 1. As a result, by calculating $\frac{b}{a}$, the NV$^-$/NV$^0$ PL ratio can be evaluated.

Carefully observing the trends, an increase in the ratio occurs at lighter oxidation conditions (i.e., at 450 °C as the process duration increases or 475 °C for low process time), consistently with the increase in NV$^-$ centers due to surface oxidation.

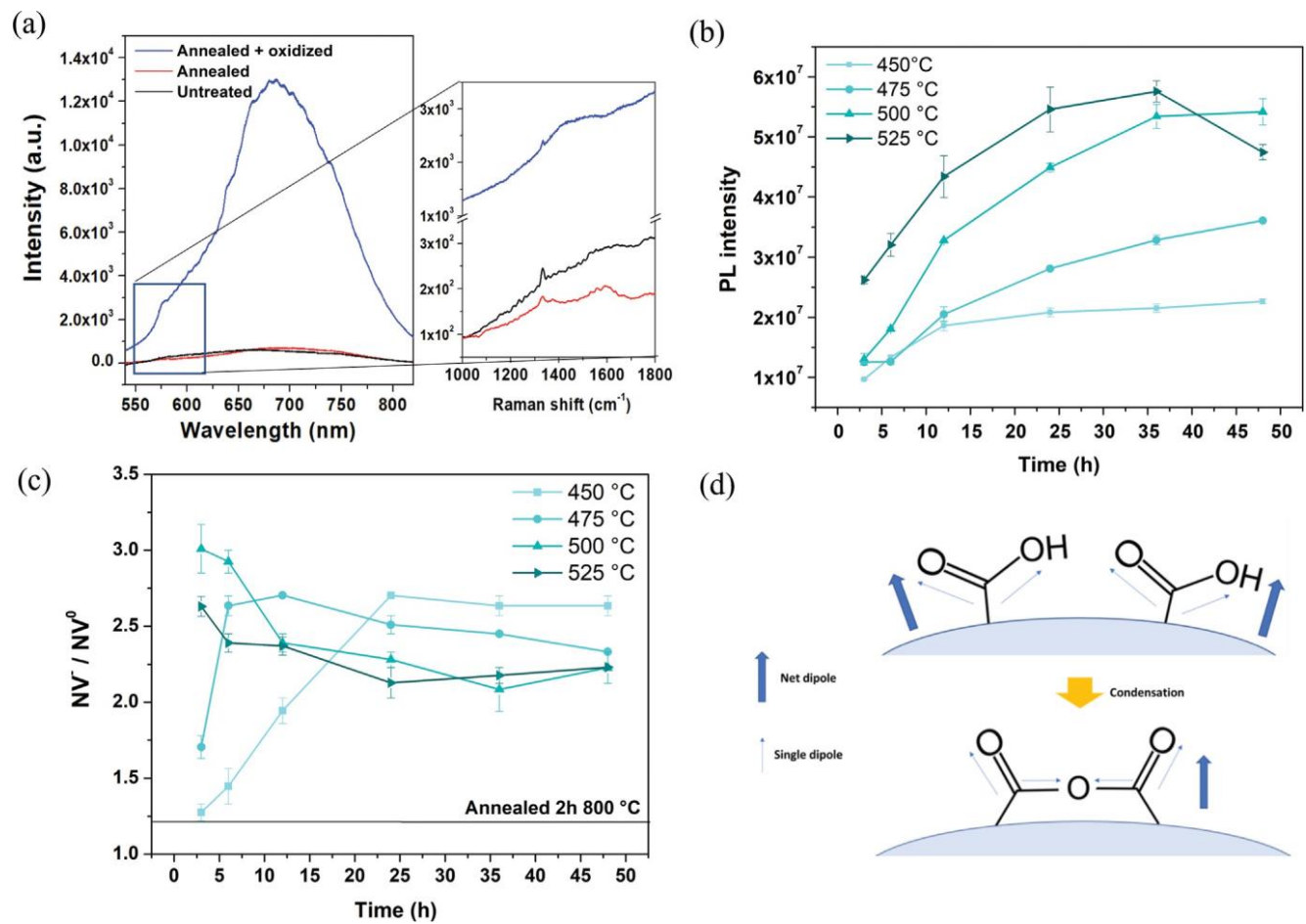

**Figure 2.** a) Raman/PL spectra of untreated, annealed and annealed + oxidized NDs; normalized spectra are also reported for better visualization in Figure S2a (Supporting Information); b) Integral of NV centers PL intensity and c) NV$^-$/NV$^0$ ratio as a function of oxidation temperature and time. d) Schematics of the dipole moment of surface C═O groups. As can be evinced in this example regarding the condensation of carboxyl acids into anhydrides, the formation of such complex structures can reduce the net dipole moment. Similar phenomena can occur in the formation of lactones or esters.

More precisely, the effect is associated with the positive dipole moment formed between carbon and oxygen atoms, which in the proximity of the surface determines the bending of the level structures, thus favoring NV$^-$ centers. Surprisingly, intermediate oxidations seem to play an opposite role by slightly decreasing this ratio. Although the reason for this phenomenon is not completely clear, a possible explanation can be found in the progressive formation of different varieties of oxygenated moieties. More precisely, it has been shown before that exactly in correspondence of intermediate oxidation temperatures and process times, the transition from isolated C═O surface groups to more complex and interconnected structures, such as anhydrides and lactones, might occur. As exemplified in Figure 2d, due to the possible presence of opposite dipole moments, this kind of chemical groups can present a net dipole that is lower with respect to that of single carboxylic acids or carbonyls present in DRIFT spectra of mildly oxidized NDs, potentially explaining the partial decrease in the NV$^-$/NV$^0$ ratio. Other alternative explanations could be attributed to the gradual removal of surface contaminations which might have a quenching effect more pronounced on NV$^0$ centers with respect to NV$^-$, or the water interaction with the sur-face chemical bonds which might also slightly reduce the dipole moment of surface oxygenated groups. In contrast, the gradual etching of surface graphite, as the oxidation level increased, might have reduced the thermal heating of the NDs under laser irradiation during the PL analysis,[35] thus decreasing the ratio.[39]

Finally, at 525 °C for oxidation time > 24 h, the ratio slightly increases again, probably due to the reduction in the crystals dimension. Consequently, the larger surface of the oxidized area might favor a further increase in NV$^-$ centers. In addition, the strong increase in C-O species observed in DRIFT spectra at such high temperatures and process durations, by promoting the general surface oxidation status, could have determined a further increase in NV$^-$ centers.

## 2.2. Ion Irradiation

In **Figure 3a** the PL spectra of NDs before and following the ion irradiation (see Experimental Section for details) in the exemplary case of a fluence F = 4.4 × 10$^{16}$ cm$^{-2}$ are shown: after irradiation, together with a significant decrease in the PL intensity, the GR1 band appears ≈745 nm, which is a spectral feature associated to individual vacancies in the diamond lattice commonly expected in implanted diamond preceding HTTA.[1] On the contrary, the PL spectrum acquired following the thermal annealing corroborates the coupling of N impurities with the newly generated vacancies by showing a strongly enhanced intensity, the disappearance of the GR1 band and a better outlining of the ZPL of the NV⁰ and NV⁻ centers. Figure 3b shows the fluorescence intensity from the PL spectra of the samples (integrated from 5G5 to 780 nm) as a function of the delivered fluence, both before (red data) and after (blue data) HTTA. Grey line represents the integrated fluorescence from unirradiated annealed + oxidized NDs. Data are calculated from the average of the integrals of 3-5 spectra acquired for each sample and the error bar is the maximum semidispersion. Due to inhomogeneities of the spatial distribution of the 5 × 5 mm² ion beam during the irradi-ation, a 20% spatial variation of the fluence values should be also considered.

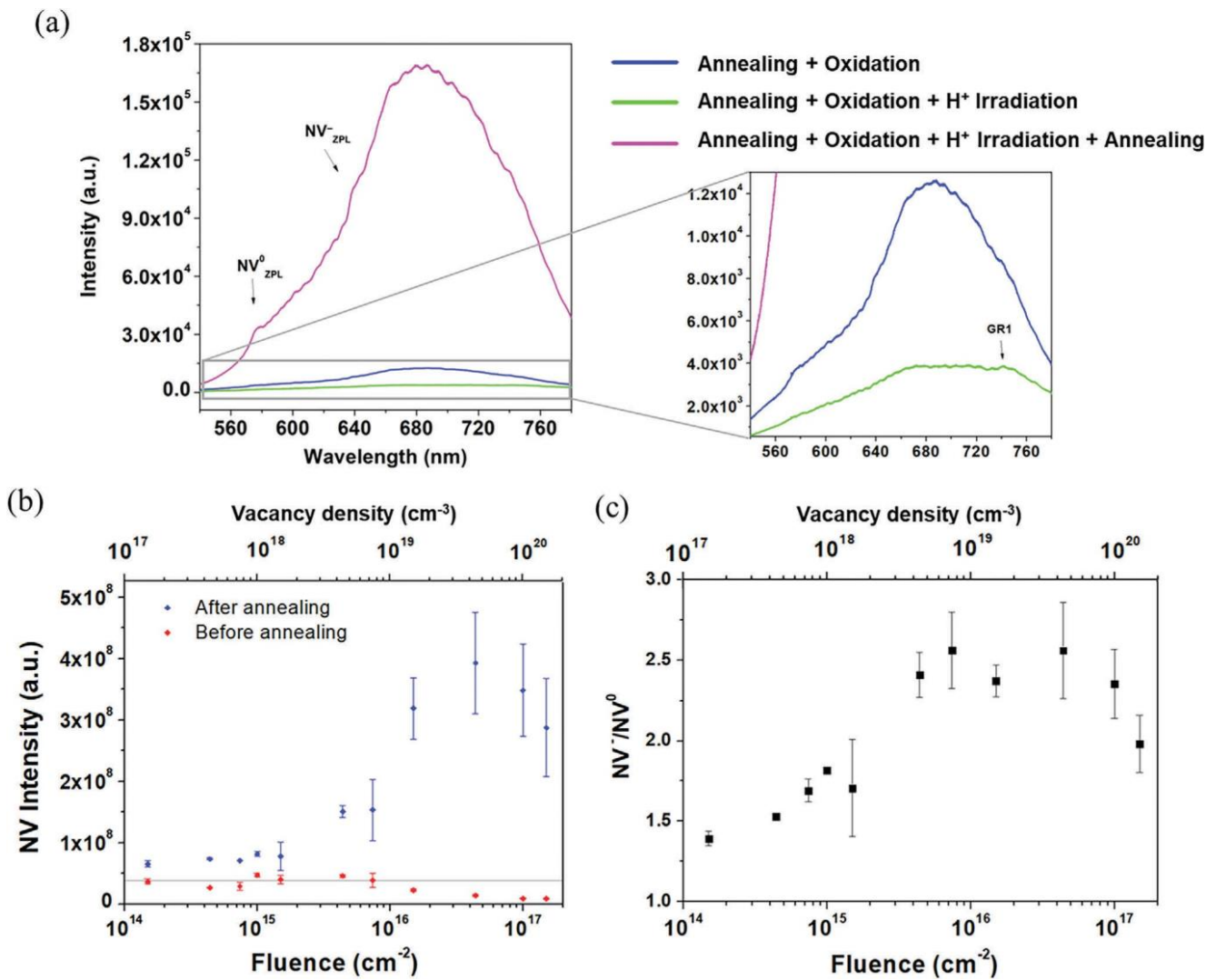

**Figure 3.** a) spectra of NDs before and following ion implantation (F = 4.4 × 10¹⁶ cm⁻²) and HTTA; normalized spectra are also reported for better visualization in Figure S2b (Supporting Information); b) integrated NV PL intensity of NDs as a function of the delivered fluence (or the damage density) before and after HTTA; gray line represents the PL intensity before ion implantation; c) NV⁻/NV⁰ ratio as a function of the fluence (or the damage density).

Before the HTTA process, at low fluences there are no significant variations in PL intensity, while it starts decreasing at higher fluences. This effect can be explained by the quenching effect due to the new vacancies which are randomly introduced into the lattice by the proton irradiation.[40] After HTTA, the PL increases at all damage levels, reaching the highest intensity at the fluence of 4.4 × 10¹⁶ cm⁻², which corresponds to a vacancy density of ≈4 × 10¹⁹ cm⁻³. The latter value is compatible with the damage level at which the maximum fluorescence from NV centers was obtained following He implantation in a previous work for bulk diamond.[41] Once this value is overcome, the PL intensity starts decreasing.

Further analysis is carried out to evaluate the ratio of the PL contribution of the two charge states of the NV centers. Figure 3c shows the trend of NV⁻/NV⁰ PL ratio as a function of the vacancy density. After a general increase at lower fluences, at higher damage levels a flattening and finally a decrease (for $1.5 \times 10^{17}$ cm$^{-2}$) of the trend is observed. According to a previous work,[41] by increasing the fluence of implantation, a decrease of the NV$^-$/NV$^0$ ratio is expected. In fact, NV centers becomes negatively charged because of the charge transfer from the substitutional nitrogen atoms, which act as donors in the crystal. The higher the implantation fluence, the higher will be the number of newly created NV centers; as a result, the number of isolated N atoms acting as electrons donors is reduced, resulting in the increase of the relative number of NV$^-$ centers. However, results in[41] were referred only to bulk diamond and the decrease of the ratio was observed only at higher damage levels with respect to the case under exam, starting around a damage density of $10^{20}$ cm$^{-3}$, which approximately correspond to the maximum fluences reached in this work. Nonetheless, while the initial decrease of the ratio in correspondence of the latter point is well explainable as mentioned above, the increase at lower densities is still unclear, but was already observed in NDs in another work,[10] whose authors ascribed the effect to the influence of the oxidized surface. However, the measurements of Figure 3c are conducted following HTTA, thus being essentially depleted of oxygen-containing species and suggesting the need of further future investigation to solve this issue.

### 2.2.1. Modeling of NV Centers Formation via Ion Irradiation

Considering the atomic density of diamond ($\rho_d = 1.77 \times 10^{23}$ cm$^{-3}$) and a nominal concentration of nitrogen impurities of the order of 100 ppm (typical of High-Pressure High-Temperature – HPHT – diamond), the volume density of nitrogen impurities can be estimated as $\rho_0 \approx 1.77 \times 10^{19}$ cm$^{-3}$. Correspondingly, the density of ion-induced vacancies ($\rho_{vac}$) for the fluence value associated to the maximum measured PL intensity (i.e., $F \approx 4 \times 10^{16}$ cm$^{-2}$) can be estimated as $\rho_{vac} \approx 4 \times 10^{19}$ cm$^{-3}$, and thus is comparable with the above-mentioned density of nitrogen atoms. This evidence suggests at first analysis that the explanation for the maximum PL intensity observed at this fluence, and the following decrease is related to the saturation of N impurities with vacancies. Nonetheless, the NV center optical emission is also limited by the actual density of defects, due to phenomena of PL quenching (i.e., resonant non-radiative energy transfer) between the NV emitters and surrounding defect complexes.

In the formation of NV centers, different mechanisms must be taken into account. To describe the phenomenon, a differential equation can be formulated taking into consideration that an infinitesimal variation of vacancy density $d\rho_{vac}$ will proportionally impact on the infinitesimal variation in the density of created NV centers $d\rho_{NV}$. In this formulation, $d\rho_{vac}$ is evaluated as the product of the numerically predicted linear vacancy density per ion ($\lambda = 1 \times 10^3$ vacancies ion$^{-1}$ cm$^{-1}$, as estimated from Figure 8d) with the infinitesimal variation in the delivered fluence $dF$). In addition, $d\rho_{NV}$ is proportional to the concurrent density of single substitutional nitrogen impurities $\rho_N$. A key parameter that must be considered affecting the nitrogen/vacancy coupling probability is the diffusion volume of the vacancies during the annealing process: the larger is such volume, the higher is the probability of approaching a nitrogen atom and forming an NV complex. NDs employed in this work are annealed for 2 h at 800 °C after the ion irradiation. By applying the model described in,[42] it can be calculated that, following thermal annealing in these conditions, vacancies can migrate up to approximately d $\approx$ 160 nm (Equation (3)):

$$d \sim \sqrt{D_0 \, t \, e^{-\frac{E_a}{kT}}} \tag{3}$$

where $D_0 = 3.G \times 10^{-6}$ cm$^2$ s$^{-1}$ is the diffusion coefficient, $E_a = 1.7$ eV is the activation energy for vacancy diffusion, k is the Boltzmann constant, T is the annealing temperature and t the annealing time. Being d higher than the median diameter of a single NDs, the vacancy diffusion volume was set to the NDs volume (i.e., considering 55 nm median diameter, V $= 8.7 \times 10^{-17}$ cm$^3$). Finally, an empirical parameter $\eta$ is introduced accounting to the efficiency in the agglomeration of NV centers per unity density of the corresponding defects (vacancies, substitutional nitrogen), thus obtaining Equation (4):

$$d\rho_{NV} = \eta \rho_N \lambda dF = \eta \rho_N V d\rho_{vac} \tag{4}$$

It should be noted that a direct proportionality between $d\rho_{NV}$ and $d\rho_{vac}$ is considered, under the assumption that the efficiency in the agglomeration of NV centers ($\eta$) does not depend on the vacancy concentration in the damage density under exam, as confirmed in.[21]

The 800 °C annealing temperature is not sufficient to de-aggregate the NV complexes,[1] therefore the irradiation + annealing process results in a steady increase of the NV centers concentration. In turn, the concurrent density of single nitrogen atoms $\rho_N$ can be expressed as the difference between the total density of nitrogen impurities $\rho_{N_{tot}}$ and the density of NV centers $\rho_{NV}$.

As a result, the infinitesimal increase of vacancy density in correspondence of an infinitesimal increase in delivered fluence can be expressed as follows (Equation (5)):

$$d\rho_{NV} = \eta \left( \rho_{N_{tot}} - \rho_{NV} \right) V d\rho_{vac} \tag{5}$$

To separate the contributions of the two charge states, the differential equation can be split for NV$^0$ and NV$^-$ color centers, as follows (Equations (G)–(8)):

$$d\rho_{NV} = d\rho_{NV^0} + d\rho_{NV^-} \tag{G}$$

$$d\rho_{NV^0} = \eta_0 \left( \rho_{N_{tot}} - \rho_{NV^0} - \rho_{NV^-} \right) V d\rho_{vac} \tag{7}$$

$$d\rho_{NV^-} = \eta_- \left( \rho_{N_{tot}} - \rho_{NV^0} - \rho_{NV^-} \right) V d\rho_{vac} \tag{8}$$

where $\eta_0$ and $\eta_-$ are the creation efficiencies for NV$^0$ and NV$^-$ centers, respectively. Assuming that the concentration of NV complex in different charge states is negligible, we can infer that $\eta_0 + \eta_- = \eta$.

We defined $\beta(\rho_{NV^0}) = \frac{\rho_{NV^0}}{\rho_{N_{tot}}}$ and $\beta(\rho_{NV^-}) = \frac{\rho_{NV^-}}{\rho_{N_{tot}}}$ as the fractions of nitrogen atoms respectively involved in the formation of NV$^0$ and NV$^-$ centers. By integrating the system of differential Equations ([7]) and ([8]), the following expressions are obtained (Equations ([9]) and ([10])):

$$\alpha\left(\rho_{vac}\right) = \frac{\eta_0\left(\beta_0 + \alpha_0 - 1\right)e^{-(\eta_0 + \eta_-)V\rho_{vac}} - \eta_0\beta_0 + \eta_-\alpha_0 + \eta_0}{\eta_0 + \eta_-} \tag{9}$$

$$\beta\left(\rho_{vac}\right) = \frac{\eta_-\left(\beta_0 + \alpha_0 - 1\right)e^{-(\eta_0 + \eta_-)V\rho_{vac}} + \eta_0\beta_0 - \eta_-\alpha_0 + \eta_-}{\eta_0 + \eta_-} \tag{10}$$

where $\alpha_0$ and $\beta_0$ are $\alpha$ and $\beta$ values in the unirradiated sample ($\rho_{vac} = 0$), i.e., they represent fraction of nitrogen atoms involved in the formation of native $NV^0$ and $NV^-$ centers.

The intensity of the acquired PL can be assumed to be proportional to $\alpha$ and $\beta$, but the mere density of emitters is not sufficient to fully account in an exhaustive model of the PL emission intensity. In a previous work, Gatto Monticone et al.[40] showed, in the case of MeV proton irradiation in bulk diamond, that upon the overcoming of a damage density threshold of $1 \times 10^{19}$ cm$^{-3}$, a decrease in the $NV^0$ and $NV^-$ centers lifetime occurs. This is explained in terms of a decrease in their PL emission quantum efficiency, and it is indicative of the key role played by isolated vacancy and interstitial defects in determining a quenching in the PL emission intensity from NV emitters. In these defect-induced PL-quenching dynamics, the contribution of more complex defects cannot be ruled out, but in the present model only the role of isolated defects will be considered. In this approximation, the dependence of the PL emission quantum efficiency from the vacancy defect concentration is thus accounted in the expression of the $\varepsilon(\rho_{vac})$ function. The $\varepsilon(\rho_{vac})$ trend was numerically extrapolated from,[40] both for $NV^0$ and $NV^-$ color centers (see Figure S7 and Equations S4 and S5, Supporting Information).

By taking the PL emission quantum efficiency into due account, the PL emission intensity for $NV^0$ and $NV^-$ centers can be described as (Equations ([11]) and ([12])):

$$PL_{NV0}\left(\rho_{vac}\right) = c_a \cdot \alpha\left(\rho_{vac}\right) \cdot \varepsilon_{NV0}\left(\rho_{vac}\right) \tag{11}$$

$$PL_{NV-}\left(\rho_{vac}\right) = c_b \cdot \beta\left(\rho_{vac}\right) \cdot \varepsilon_{NV-}\left(\rho_{vac}\right) \tag{12}$$

where $c_a$ and $c_b$ are proportionality constants introduced to rescale the function intensity with the instrumental output (i.e.,: accounting for average PL collection efficiency, excitation power and wavelength influence, etc.).

With the purpose of avoiding over-fitting, some of the above-introduced parameters can be fixed based on the results of previous works. By analyzing a work of Pezzagna et al.,[21] an indicative estimation of the parameter $\eta$ = G.G × 10$^{-4}$ was evaluated (see Section S5, Supporting Information). Fixing this parameter allows to decrease the number of free parameters in our model (Equation ([13])):

$$\eta_0 = G.G \times 10^{-4} - \eta_- \tag{13}$$

A similar relation between $\alpha_0$ and $\beta_0$ can be set prior to data fitting, by analyzing the spectrum acquired from the unirradiated sample. Such PL spectrum intensity at F = 0 cm$^{-2}$ ($PL^{(0)}$) can be split in the spectral contribution of the two classes of emitters, i.e., $PL^{(0)}_{NV^0} = a^{(0)} \cdot PL^{(0)}$ and $PL^{(0)}_{NV^-} = b^{(0)} \cdot PL^{(0)}$, by adopting the same method that was previously employed to estimate the $NV^-$/$NV^0$ ratio, thus obtaining an estimation of $\frac{a^{(0)}}{b^{(0)}}$ = 0.73.

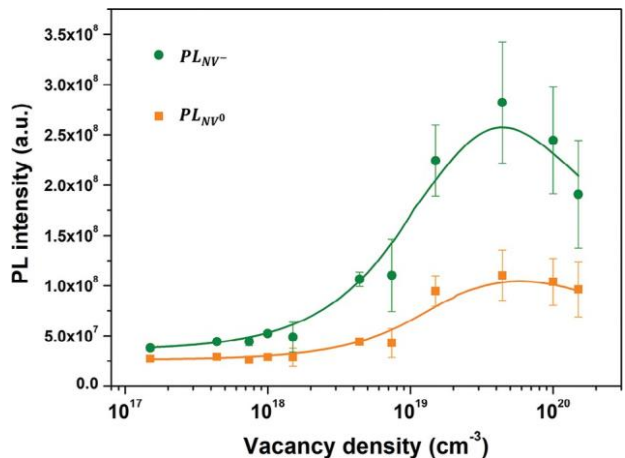

**Figure 4.** $NV^0$ (orange) and $NV^-$ (green) PL data and fit with Equations ([11]) and ([12]).

$\alpha_0$ and $\beta_0$ can be linked to the observed PL as follows (Equations ([14]) and ([15])):

$$PL^{(0)}_{NV^0} = a^{(0)}\,PL^{(0)} = j\,\alpha_0 \tag{14}$$

$$PL^{(0)}_{NV^-} = b^{(0)}\,PL^{(0)} = k\,\beta_0 \tag{15}$$

where $j$ and $k$ are defined as proportionality parameters linking the PL emission intensities from the two types of emitters (i.e., $NV^0$ and $NV^-$) to their corresponding concentrations. By dividing the two equations, we obtain (Equation ([1G])):

$$\alpha_0 = \frac{a^{(0)}}{b^{(0)}} \cdot \frac{k}{j}\beta_0 \tag{1G}$$

where $\frac{k}{j}$ ratio identifies the ratio between the PL emission intensities of equally concentrated $NV^-$ and $NV^0$ centers, at F = 0 cm$^{-2}$. Besides the difference in terms of investigated samples type, in first approximation a value of $\frac{k}{j}$ = (2.5 ± 0.5) can be extrapolated from,[38] where PL characterization of a CVD grown diamond sample was performed using a low power (<20 kW cm$^{-2}$) 532 nm laser excitation, consistently with the present work. By substituting the values ($\frac{a^{(0)}}{b^{(0)}}$ = 0.73 and $\frac{k}{j}$ = 2.5) in Equation ([1G]), the following relation is obtained (Equation ([17])):

$$\alpha_0 = 1.82\,\beta_0 \tag{17}$$

This relation, together with the one obtained in Equation ([13]), are introduced in Equations ([9]) and ([10]), thus obtaining two functions with a reduced number of free parameters (namely: $c_a$, $c_b$, $\beta_0$ and $\eta_-$). Total NV intensities from Figure 3b are split in $NV^0$ and $NV^-$ components (see Section S3, Supporting Information), obtaining the plot of **Figure 4**. $NV^0$ and $NV^-$ PL intensity data are fitted with the functions obtained in Equations ([11]) and ([12]), respectively. Fit values are reported in **Table 1**.

The obtained values of $\eta_-$ and $\beta_0$ appear compatible within a 10% of error, which is reasonable to be considered in light of the uncertainties associated with the imposed assumptions and approximations at the basis of the model. Thus, average values of $\eta_-$ and $\beta_0$ were obtained from the fits, and by substituting them in Equations (13) and (17), $\eta_0$ and $\alpha_0$ were calculated and reported in **Table 2**.

**Table 1.** Results of parameters from fitted data of $NV^0$ and $NV^-$ PL with Equations (11) and (12), respectively. 10% error is considered on each parameter.

|  | Fit NV0 | Fit NV− |
|---|---|---|
| $c_a$ | $1.1 \times 10^7$ | – |
| $c_b$ | – | $2.0 \times 10^7$ |
| $\eta_-$ | $5.1 \times 10^{-4}$ | $4.5 \times 10^{-4}$ |
| $\beta_0$ | 0.061 | 0.064 |

As shown in Figure 4, the fitting functions suitably interpolate both experimental PL intensity trends as a function of the vacancy density, thus corroborating the validity of the adopted model. It is worth remarking that a higher value of $\eta_-$ with respect to $\eta_0$ is coherent with the commonly observed evidence that NV centers are more easily created in their negative charge state in Ib-type HPHT diamond samples characterized by a n-type doping due to the relatively high concentration of substitutional nitrogen (with respect, for example, to type IIa samples).[41] The $\alpha_0$ and $\beta_0$ parameters provide an estimation of the starting portion of single substitutional nitrogen impurities involved in the formation of $NV^0$ and $NV^-$ centers, respectively. By approximately assuming 100 ppm concentration of N impurities (typical of HPHT diamonds), the starting densities of $NV^0$ and $NV^-$ centers can be estimated, resulting in $\rho_{NV0}(F = 0) \approx 2.0 \times 10^{18}$ cm$^{-3}$ and $\rho_{NV-}(F = 0) \approx 1.1 \times 10^{18}$ cm$^{-3}$, respectively, corresponding approximately to 200–300 NV centers per NDs. Similarly, this can be calculated when considering the vacancy density at which the maximum fluorescence was measured, thus obtaining $\sim 1400$–1500 NV centers / NDs, allowing to conclude that the process of optimization of the PL emission intensity was obtained upon a ×5-G increase in the NV center concentration.

After substituting the obtained parameters in Equations (9) and (10), the $\alpha(\rho_{vac})$ and $\beta(\rho_{vac})$ functions can be plotted (see **Figure 5**), showing how the saturation of almost all N impurities is achieved at vacancy densities of $\approx 10^{20}$ cm$^{-3}$. Considering an approximate density of native substitutional N impurities of $\approx 10^{19}$ cm$^{-3}$, it can be evinced that, in the case study of 55 nm diameter NDs in "nitrogen saturation" conditions, 9 out of 10 ion-generated vacancies remain uncoupled from N impurities, arguably upon their aggregation in larger defective complexes and/or their migration to the nanoparticle surface during the thermal annealing process. The saturation of N impurities with vacancies occurs when an irradiation fluence of $\approx 10^{17}$ cm$^{-2}$ is delivered. Such value is slightly larger than the fluence corresponding to the maximization of the PL emission intensity (namely, F = 4–5 × $10^{16}$ cm$^{-2}$), due to the quenching phenomena occurring at damage densities higher than $1 \times 10^{19}$ vac cm$^{-3}$.

**Table 2.** Calculated parameters from results reported in Table 1 and Equations (13) and (17).

| | |
|---|---|
| $\alpha_0$ | $0.114 \pm 0.011$ |
| $\beta_0$ | $0.063 \pm 0.006$ |
| $\eta_0$ | $(1.8 \pm 0.2) \times 10^{-4}$ |
| $\eta_-$ | $(4.8 \pm 0.5) \times 10^{-4}$ |

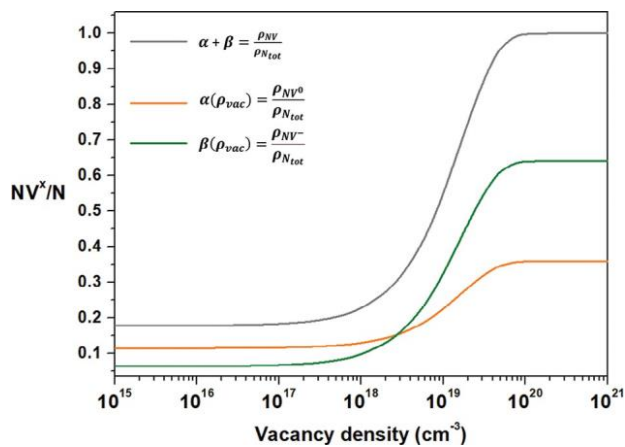

**Figure 5.** Plot of the ratios of $NV^0$ (orange plot) and $NV^-$ (green plot) centers densities with respect to the density of N impurities. Total NV/N ratio is also plotted in Gray, reaching saturation $\approx 10^{20}$ cm$^{-3}$ of vacancy density.

Moreover, this phenomenon can explain the reason behind the observed increase of more than 1 order of magnitude of the PL intensity with only 5–G times increase in the total number of generated NV centers. Indeed, the most significant increase of NV centers is associated with the creation of the NV centers in negative charged state, which are characterized by a significantly larger PL emission quantum efficiency with respect to the NV centers in the neutral charge state.

## 2.3. Processing Optimization

Following the characterizations presented in the previous sections, NDs annealed and oxidized for 3G h at 500 °C are selected as they reach very high fluorescence values while avoiding excessive degradation that might occur at 525 °C, and irradiated with a fluence of $4.4 \times 10^{16}$ cm$^{-2}$ (corresponding to the fluence that maximized the PL intensity).

Following the irradiation, the samples are processed and characterized with Raman/PL spectroscopy. First, the ideal time of re-annealing after ion irradiation was investigated. To this scope, annealed + oxidized samples are annealed in 3 successive times in Ar flow for 2 h at 800 °C. Figure S9 (Supporting Information) summarizes the results in terms of integrated PL intensity. It can be evinced that, after the second re-annealing, the fluorescence increase is no longer observed, allowing to conclude that a thermal process 4 h long is sufficient to obtain the maximum NV centers formation, while further prolonging the treatment could even turn out counterproductive (mild graphitization can occur). Following the 4 h HTTA, PL intensity shows an increase greater than 1 order of magnitude (**Figure 6**). Samples are then treated with repeated air oxidations for 12 h at 500 °C, showing only a very weak enhancement and even a slight detrimental effect, as the process gets too aggressive.

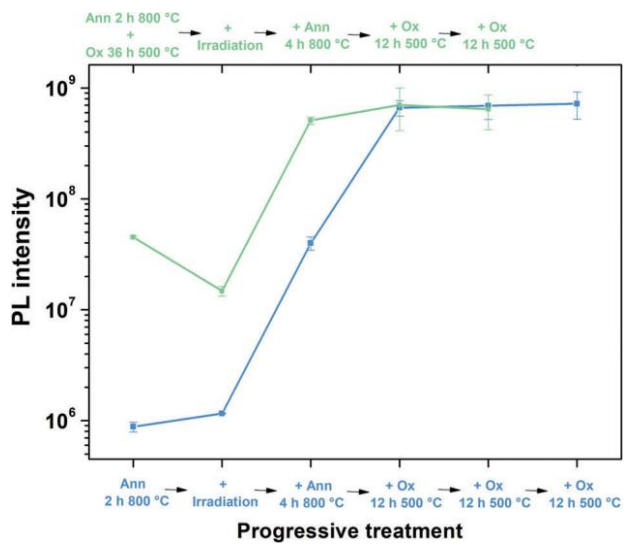

**Figure 6.** PL intensity at progressive treatments of irradiation, HTTA (ann) and oxidation (ox) for both previously oxidized (green) and not-oxidized (blue) NDs.

With this protocol, the integrated PL results increased of ≈3 orders of magnitude with respect to unprocessed NDs.

## 2.4. PL Decay Measurements

To verify the effectiveness of the performed purification processes in reducing the optical quenching by defective phases, PL decay measurements are carried out. The presence of a defective system or an impurity located in proximity of a fluorescent center (generally referred as a "quencher" of the optical transition) determines the onset of a new non-radiative decay channel for the center. The different resonant coupling strengths for randomly located quenchers around the luminescent center result in a deformed exponential decay rate[40,43,44] (Equation 18):

$$I\left(t\right)=\frac{N_{0}}{\tau}\,\mathrm{e}^{-\frac{t}{\tau}}\mathrm{e}^{-k\left(\frac{t}{\tau}\right)^{c}} \qquad (18)$$

where $k$ is an a-dimensional parameter dependent from the quenchers concentration and the intensity of the resonant coupling, $N_0$ is the PL intensity at t = 0 s, while the value of $c$ depends on the nature of the resonant coupling (e.g.,: $c$ = ½ for dipole-dipole interaction, $c$ = ⅜ for dipole-quadrupole interaction).

Equation (18) is employed to fit PL decay data acquired from mildly (i.e., G h at 450 °C processing), intermediately (i.e., G h at 500 °C processing) and highly (i.e., 3G h at 500 °C processing) oxidized NDs. Samples are prepared by dispersing and sonicating NDs in isopropyl alcohol solutions (≈0.1 mg ml⁻¹). A drop of the resulting solutions is deposited on a silicon wafer and dried. The PL decay measurements are repeated on G–8 spots for each sample under analysis, following the protocol reported in Experimental Section ("PL decay measurements set-up").

Exemplary PL decay curves are plotted in **Figure 7a**. To extrapolate the values of the $k$ parameter, PL decay data are fitted by leaving $N_0$ and $k$ as free parameters, while, due to better fitting accuracy, $c$ is set to ½ and the value of $\tau$ is imposed equal to 80 ns. Although atypical, similarly high values of lifetime have been already reported in literature,[45] motivated by Purcell effect which determines higher lifetime in NDs with respect to bulk diamond.[46] The difference with respect to typical NV⁻ $\tau$ values (18–25 ns[30,47]) is justified by the different decay rate model employed in this work, which separately accounts for the contribution of the quenching effect described by $k$ parameter.

As reported in Figure 7b, clear evidence results on the reduction of the $k$ value when passing from very mild to intermediate oxidation conditions, which can be associated with a significant reduction of the non-radiative quenching level following oxidation. A very slight increase in $k$ value, although not statistically significant, is then observed for the highest oxidation level. Interestingly, it can be noted that similar trend was observed in the PL NV⁻/NV⁰ ratio, where the highest value was indeed observed in the intermediate oxidation conditions here considered, thus corroborating the better PL performance of NV⁻ centers at this oxidation level.

Overall, the results suggest that a significant role in the enhancement of the PL intensity reported following oxidation in Section 2.1 is played by the removal of defective quenching phases surrounding NDs core.

The PL decay analysis is also performed for the annealed + oxidized (i.e., 3 h at 525 °C) samples irradiated with protons at 4 × 10¹⁴, 1.5 × 10¹⁶, and 1.5 × 10¹⁷ cm⁻² fluences, following further thermal annealing for 2 h at 800 °C. Results are reported in Figure 7c,d, showing similar quenching level in the cases of NDs irradiated at 4 × 10¹⁴ and 1.5 × 10¹⁶ cm⁻². The similar PL decay profile among the two above-mentioned samples, together with the strong enhancement in PL intensity shown in Section 2.2 at 1.5 × 10¹⁶ cm⁻², support the effectiveness of the ion irradiation process in increasing the amount of NV centers without determining potential quenching phenomena associated with the introduction of other defects. Different result is obtained at the highest explored fluence of 1.5 × 10¹⁷ cm⁻², when a significant increase in the quenching level (i.e., higher k) occurs. This result can be interpreted with the creation of a high density of PL-quenching defects due to the ion-induced damaging of NDs and is fully consistent with the decrease in PL evidenced at this fluence reported in Figure 3b.

## 3. Conclusion

In the present work, NDs are systematically processed by means of thermal oxidation treatments and proton beam irradiation to tune their optical and surface chemistry properties. In general, higher oxidation levels result correlated with a higher number of oxygen-containing chemical groups and enhanced NV fluorescence. Mild oxidations determine the appearance of mainly carboxylic acids and anhydrides, while stronger processes are associated with a further increase of these species together with the supposed appearance of aldehydes, lactones, and ketones. At highly aggressive processing conditions, a strong increase of mainly C-O groups is observed. All processes are associated with an increase of surface hydrophilicity, which is essential for biomedical applications due the consequent better dispersibility and stability of NDs in water solution.[18] Regarding PL analysis, a tendency to a saturation is observed at the highest oxidation degrees, due to the almost complete removal of surface quenching phases, while excessive oxidation resulted in slight decrease in fluorescence, as also the diamond core phase is etched. The effectiveness of the oxidation treatments in the removal of surface quenching phases is also confirmed by PL decay measurements performed on selected samples. $NV^-/NV^0$ ratio shows the highest value in the mildly oxidized samples, thus suggesting them as preferable when a high concentration of $NV^-$, for applications such as thermometry/magnetometry, is required A 2 MeV proton beam irradiation proves effective in drastically increasing the PL emission of NV centers from NDs. In the explored values, a fluence of $4.4 \times 10^{16}$ cm$^{-2}$, corresponding to a damage level of $4 \times 10^{19}$ vacancies cm$^{-3}$, provides the highest fluorescence intensity, by determining ≈1 order of magnitude increase with respect to unirradiated samples. A predictive model is elaborated to describe the trend of PL of both $NV^0$ and $NV^-$ centers as a function of the vacancy density, by considering the NV centers formation mechanism and the quantum efficiency as a function of the damage density. The model properly describes experimental data, providing a value of the $NV^-$ creation efficiency $\eta^-$ higher than that of $NV^0$ $\eta^0$, suggesting an easier creation of $NV^-$ centers with respect to $NV^0$ centers, probably due to a significant number of charges available on the surface. The results of the model fit allow also to estimate the relative amount of $NV^-$ and $NV^0$ with respect to N impurities in the unirradiated samples, which resulted indicatively equal to 6% and 11%, respectively. Based on the developed model, the variation in the initial distributions of NV center charge states was described, predicting their inversion upon ion irradiation. In addition, it can be inferred that almost all N impurities are involved in NV center formation ≈10$^{20}$ cm$^{-3}$ of vacancy density, which is 10 times higher than the approximate value of the density of N impurities. PL decay measurements performed at different fluences demonstrated that at fluences below $1.5 \times 10^{16}$ cm$^{-2}$ no significant variation occurs in the concentration of defective quenchers, thus supporting the association of the increase in PL with the effective increase in NV centers concentration. Conversely, an increase in the quenching level is observed at the highest fluence of $1.5 \times 10^{17}$ cm$^{-2}$, consistently with the formation of a high density of PL-quenching defects in the NDs, which is confirmed by the observed decrease in PL intensity.

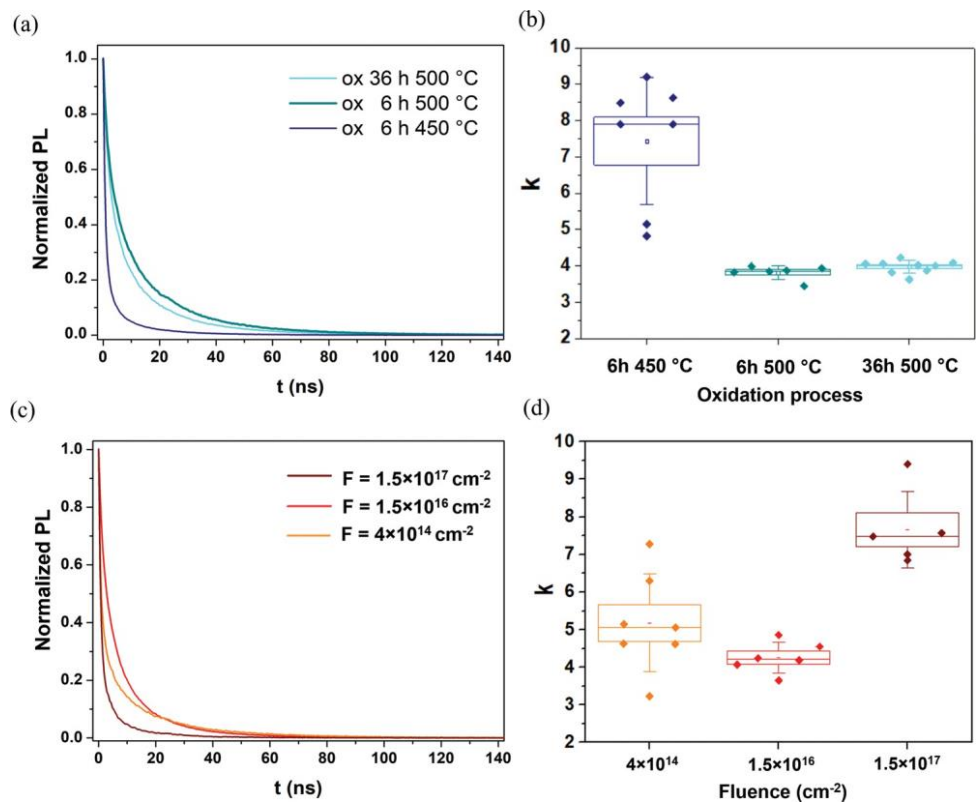

**Figure 7.** a) Normalized PL decay measurements and b) $k$ values obtained from NDs at different oxidation levels. The higher $k$ value observed at mild oxidation is associated with higher concentration of quenching defects; c) Normalized PL decay measurements and d) $k$ values obtained from NDs at different ion irradiation fluences.

Finally, it is worthwhile highlighting how, starting from untreated samples, the combination of thermal processes with ion implantation allows to reach ≈3 orders of magnitude increase in PL emission intensity, thus representing a successful protocol for PL optimization, for such applications in which general fluorescence is the most important parameter (e.g., biolabeling). Overall, the results provide fundamental insights into the connection between NDs surface chemistry, hydrophilicity, fluorescence, and NV centers charge state distribution, which are essential starting points for future NDs employment in nanomaterial and biosensing research. The developed mathematical model is successfully validated in the present case study, but its applicability is general and will represent a substantial landmark enabling to predict the outcome of ion-beam based NV centers generation in diamond, both at bulk and nanocrystalline level.

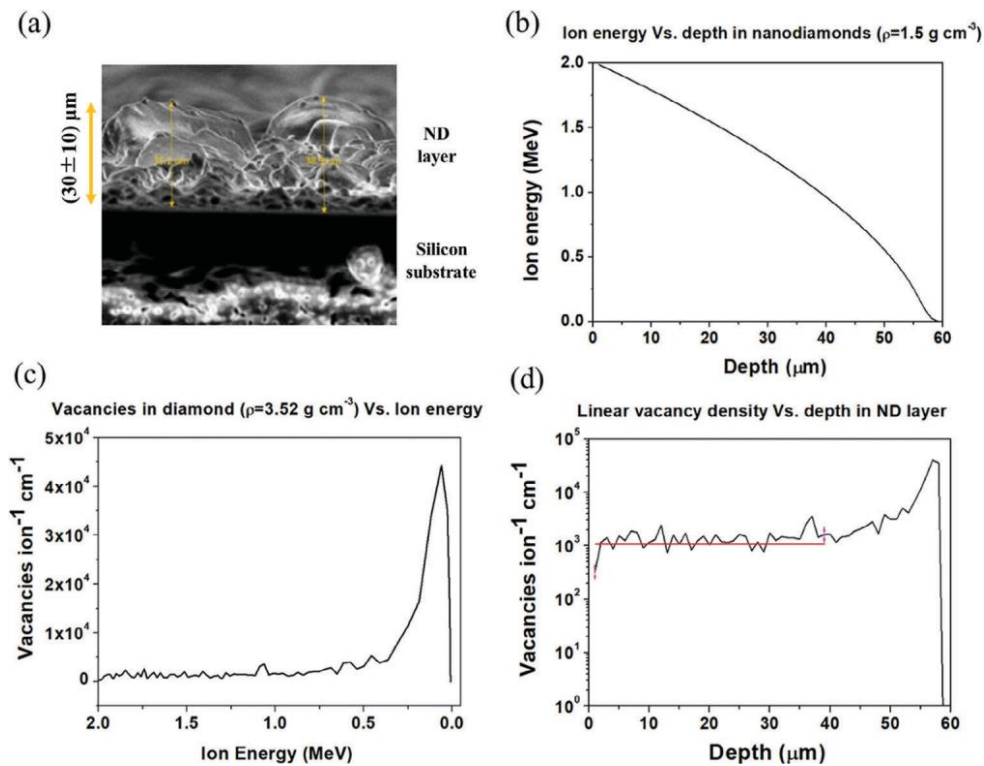

**Figure 8.** a) thickness estimation by SEM of the deposited NDs layer over the silicon substrate, resulting ~ (30 ± 10) μm; b) energy of protons as a function of the depth in the NDs layer (simulated with $\rho_{powder}$ = 1.5 g cm$^{-3}$); c) linear vacancy density per ion of protons in diamond (simulated with $\rho_{bulk}$ = 3.52 g cm$^{-3}$) as a function of their energy; d) linear vacancy density per ion realized in the NDs layer as a function of the depth, obtained by calculating the function interpolated from graph c) at the energy values of graph b).

## 4. Experimental Section

Samples employed in this work were MSY 0-0.1 HPHT monocrystalline synthetic NDs acquired from Pureon, ranging from few nm up to ≈100 nm, with a median size of 50 nm.

First, samples underwent a high temperature thermal annealing (HTTA) at 800 °C for 2 h in nitrogen flow. The HTTA process was conducted to reorganize the disordered sp²/sp³ phases[48] and graphitize the amorphous carbon outer layers while preserving the diamond phase. In addition, the lack of oxygen induces the chemical reduction of the surface of the nanocrystals. The setting of an optimal annealing temperature was of fundamental importance to avoid undesired structural and chemical degradation of the NDs during the process. On one hand, the nitrogen flux provides an inert environment that prevents the chemical etching of diamond that would occur above 500 °C in presence of oxygen.[36] On the other hand, despite the inert atmosphere, a temperature below 900 °C was necessary to prevent the samples from graphitization[34,49] and the conversion to "carbon onion" structures.[50] Therefore, the HTTA process was performed at 800 °C. Subsequently, to selectively remove the surface graphitic shells and to tune the surface oxidation level, systematic air thermal oxidation processes were carried at 3, 6, 12, 24, 36, and 48 h, with temperatures ranging between 450 °C and 525 °C, with 25 °C step. All the thermal treatments were performed by mean of ThermoConcept ROT 60/300/12 tubular furnace.

Following these processes, both DRIFT and Raman/PL analysis were conducted to evaluate the effect of the processes in terms of surface chemistry and optical properties, respectively.

*Ion Irradiation*: Annealed + oxidized samples (N₂ flux 2h at 800 °C + air exposure in the case of 3 h at 525 °C) were subjected to a proton irradiation process. After dispersing the NDs in isopropyl alcohol obtaining a dense suspension, a couple of drops were deposited on a silicon wafer square (≈1 × 1 cm²) and dried under an infrared lamp source. As a result, a compact ~ (30 ± 10) μm thick layer of NDs was obtained. The thickness and its variability were evaluated via SEM microscopy (**Figure 8a**). The samples were then irradiated in vacuum (≈10⁻⁸ mbar) with a broad (5 × 5 mm²) 2 MeV H⁺ ion beam at the AN2000 accelerator facility of the INFN National Laboratories of Legnaro (INFN-LNL)[51], with a beam current varying between 600-700 nA. With the purpose of studying the trend of the NDs fluorescence at different ion-damage levels, multiple depositions were irradiated at different fluences ranging from 1.5 × 10¹⁴ cm⁻² up to 1.5 × 10¹⁷ cm⁻², thus keeping sufficiently below the graphitization threshold.[52–55] The delivered fluence was estimated from the integrated charge collected by using the irradiation chamber as a Faraday cup. This allows to obtain a real time measurement of the beam current during the irradiation.

To evaluate the damage level due to the irradiation, a Monte Carlo simulation was carried out using SRIM software.[56] Although this tool works based on binary collisions approximation (BCA), the simulated output could provide a reliable estimation of the order of magnitude of the created vacancies of the case study. As simulation input, in addition to the already described irradiation parameters, the setting of the material density required more attention, since it was important to consider the powder nature of the sample. For this reason, while the density of bulk diamond was ≈3.52 g cm⁻³, due to the interstitial spaces occurring between nanocrystals, the powder density would assume a lower value. To estimate it, a 10 ml graduated cylinder was filled with the compressed powder. Once measured the contained mass, an effective density of ≈1.5 g cm⁻³ was obtained. Running the simulation with this value, the energy of the impinging protons as a function of the depth in the NDs layer was obtained (Figure 8b). Nonetheless, protons entering each single nanocrystal create a damage level equal to that which would be realized in monocrystalline diamond. The estimation of the linear vacancy density introduced in bulk diamond as a function of the ion energy was thus obtained from a simulation carried out with a target density of 3.52 g cm⁻³ (Figure 8c). By interpolating ion energy data from the latter graph with the trend of ion energy as a function of depth in the NDs layer of Figure 8b, the linear vacancy density in the NDs layer as a function of depth was plotted (Figure 8d), revealing a constant vacancy density profile until ≈40 µm, with a value of ≈1 × 10³ vacancies ion⁻¹ cm⁻¹. Therefore, when considering a NDs layer thickness lower than this value, the simulation suggests an almost flat profile along its entire volume, while the Bragg peak (maximum energy release) takes place in the silicon substrate. Finally, by multiplying the delivered fluence by the linear vacancy density, the volume vacancy density could be obtained.

After the proton irradiation, to complete the creation of NV centers by inducing the newly created vacancy to migrate and couple with the nitrogen impurities, a thermal annealing in Ar flow was carried out on all the samples for 4 h at 800 °C.

*DRIFT*: Diffuse Reflectance Infrared Fourier Transform (DRIFT) spectroscopy was performed on as-received, annealed and annealed + oxidized NDs at the different oxidation levels, with the purpose of identifying the superficial functional groups. The FT-IR spectra were recorded with a Bruker Equinox 55 FTIR spectrometer, equipped with a Mercury – Cadmium – Telluride (MCT) cryogenic detector; 64 interferograms (recorded at 2 cm⁻¹ resolution) were averaged for each spectrum. The reflectance values were successively converted in pseudo-absorbance values (i.e., A = −log R, where R is the measured reflectance).

*Raman and Photoluminescence*: Raman and photoluminescence (PL) spectra of oxidized and proton irradiated NDs were acquired with a Horiba Jobin Yvon HR800 Raman micro-spectrometer equipped with a continuous NdYAG 532 nm excitation laser, focused with a 20× air objective, and a CCD detection system with Peltier cooling system (−70 °C). This setup guarantees the collection of spectra with a spatial resolution of ≈2 µm in diameter and ≈3 µm in confocal depth. The 600 lines mm⁻¹ diffraction grating guarantees a spectral resolution of ≈3 cm⁻¹. In all measurements, the effective power incident on the sample was 17 mW and the acquisition time was set to 1 s, averaging 5 repeated acquisitions. The samples were prepared dispersing the NDs powders in isopropyl alcohol and depositing a thick and dense layer on silicon substrates. 20× objective was employed to allow the detection of an averaged signal coming from a relatively wide sample area.

*PL Decay Measurements Set-Up*: A home-built confocal microscope was employed for PL decay measurements. The setup was equipped with a NKT Photonics SuperK FIANIUM FIU-6 Supercontinuum laser, which provides the excitation pulse at a given repetition rate and intensity. A 1 MHz repetition rate was chosen to allow for the fluorescence signal to extinguish between subsequent pulses. The power was set so that pile-up-related systematic errors were negligible. A SuperK Varia tunable filter and a Thorlabs bandpass filter select the excitation wavelength (514.5 nm, FWHM = 3 nm). The probe laser was focused on the sample using a Nikon TU Plan Fluor 100×, NA = 0.90, and the same objective was used to collect the induced photoluminescence. The laser reflection and first-order Raman emission was then filtered out from the collected light with a 600 nm bandpass. PL was then split between two Excelitas SPAD-AQRH-13-FC Single Photon Avalanche Diodes (SPADs) using a 50: 50 beamsplitter. Finally, an ID Quantique ID800 Time to Digital Converter (TDC) correlates photon detections between the two SPADs. The analyzed spots were identified by checking emission spectra with a connected spectroscopy apparatus, consisting of a Teledyne SpectraPro HRS-300 monochromator paired with a Pixis:100 camera. The chosen grating has 600 mm⁻¹ density and 500 nm blaze. Spectra were background corrected. The lifetime measurements were performed by following the procedures reported.[57] Additional details regarding the measurement protocol are reported in Section S7 (Supporting Information).

## Supporting Information

Supporting Information is available from the Wiley Online Library or from the author.

## Acknowledgements

This research was supported by AURORA project funded by National Institute for Nuclear Physics (INFN), the European Union's H2020 Marie Curie ITN project LasIonDef (GA no. 956387), 20FUN02 "POLight" and "SEQUME" project funded under the EMPIR programme which is cofinanced by the Participating States and from the European Union's Horizon 2020 research and innovation programme, "Intelligent fabrication of QUANTum devices in DIAmond by Laser and Ion Irradiation" (Quant-Dia) project funded by the Italian Ministry for Instruction, University and Research within the "FISR 2019" program. This research was also supported by the coordinated research project "Sub-cellular imaging and irradiation using accelerator-based techniques" of the International Atomic Energy Agency (IAEA, CRP F11024).

## Conflict of Interest

The authors declare no conflict of interest.

## Data Availability Statement

The data that support the findings of this study are available from the corresponding author upon reasonable request.

## Keywords